\documentclass[draft]{agujournal2019}
\usepackage{url} %this package should fix any errors with URLs in refs.
\usepackage{lineno}
\usepackage[inline]{trackchanges} %for better track changes. finalnew option will compile document with changes incorporated.
\usepackage{soul}
\usepackage{color}
\usepackage{amsmath}
\usepackage{amssymb}
\usepackage[normalem]{ulem}
\usepackage{booktabs}

\draftfalse

\journalname{Geophysical Research Letters}

\begin{document}

\title{Estimating Soil Electrical Parameters in the Canadian High Arctic from Impedance Measurements of the MIST Antenna Above the Surface}

\authors{\
I. Hendricksen\affil{1}, \
R. A. Monsalve\affil{2,3,4}, \
V. Bidula\affil{1}, \
C. Altamirano\affil{4}, \
R. Bustos\affil{4}, \
C. H. Bye\affil{5}, \
H. C. Chiang\affil{1,6}, \
X. Guo\affil{2}, \
F. McGee\affil{1}, \
F. P. Mena\affil{7}, \
L. Nasu-Yu\affil{1}, \
C. Omelon\affil{8}, \
S. E. Restrepo\affil{4,9}, \
J. L. Sievers\affil{1,10}, \
L. Thomson\affil{8}, \
N. Thyagarajan\affil{11}}

\affiliation{1}{Department of Physics and Trottier Space Institute, McGill University, Montr\'eal, QC H3A 2T8, Canada}
\affiliation{2}{Space Sciences Laboratory, University of California, Berkeley, CA 94720, USA}
\affiliation{3}{School of Earth and Space Exploration, Arizona State University, Tempe, AZ 85287, USA}
\affiliation{4}{Departamento de Ingenier\'ia El\'ectrica, Universidad Cat\'olica de la Sant\'isima Concepci\'on, Alonso de Ribera 2850, Concepci\'on, Chile}
\affiliation{5}{Department of Astronomy, University of California, Berkeley, CA 94720, USA}
\affiliation{6}{School of Chemistry and Physics, University of KwaZulu-Natal, Durban, South Africa}
\affiliation{7}{National Radio Astronomy Observatory, Charlottesville, VA 22903, USA}
\affiliation{8}{Department of Geography, Queen's University, Kingston, ON K7L 3N6, Canada}
\affiliation{9}{Centro de Energ\'ia, Universidad Cat\'olica de la Sant\'isima Concepci\'on, Alonso de Ribera 2850, Concepci\'on, Chile}
\affiliation{10}{School of Mathematics, Statistics, \& Computer Science, University of KwaZulu-Natal, Durban, South Africa}
\affiliation{11}{Commonwealth Scientific and Industrial Research Organisation (CSIRO), Space \& Astronomy, P. O. Box 1130, Bentley, WA 6102, Australia}

\correspondingauthor{I. Hendricksen}{ian.hendricksen@mail.mcgill.ca}

\begin{keypoints}

\item The MIST radio cosmology experiment requires an accurate electrical characterization of its observation site in the Arctic.
\item Soil electrical parameters were estimated for MIST's observation site using the impedance of the experiment's antenna in July, 2022.
\item The estimates are consistent with a thawed active layer underlain by permafrost, and will be used to calibrate MIST sky observations.

\end{keypoints}

\begin{abstract}
The MIST experiment aims to detect the cosmological $21$-cm signal through sky observations at $25$--$125$~MHz using a wide-beam antenna. The antenna is mounted above the soil and the beam characteristics are highly dependent on the soil's electrical properties. Accurate models for the beam obtained from electromagnetic simulations are crucial for detecting the $21$-cm signal. Determining the soil properties to inform the beam simulations is therefore a very high priority for MIST. Here we report the first electrical characterization of the MIST observation site in the Canadian High Arctic, which was conducted in July, 2022. The electrical parameters were estimated using impedance measurements of the instrument's antenna, which is a very advantageous approach for MIST. Our best-fit soil model is consistent with a thawed active layer underlain by permafrost, and the parameters were estimated with a precision close to the requirements for the detection of the cosmological $21$-cm signal.

\end{abstract}

\section*{Plain Language Summary}
The MIST experiment searches for a radio signal produced in the early Universe and observes the sky with an antenna mounted above the soil. The experiment requires an accurate model of the antenna response to sky radiation but this model is highly dependent on the soil's electrical properties. This dependence makes the electrical characterization of the observation site a top priority. We report measurements of the soil electrical properties at the MIST observation site in the Canadian High Arctic conducted in July 2022. The soil was characterized using impedance measurements from the antenna itself, which is a very advantageous approach for MIST. The soil model obtained is consistent with a thawed active layer above a permafrost layer and the model parameters were estimated with a precision close to the requirements for the detection of the radio signal from the early Universe.

\section{Introduction}
\label{section_introduction}
Observations of the sky-averaged, or global, $21$-cm signal produced by neutral hydrogen gas represent one of the most promising approaches to study the first billion years of the Universe \cite{madau1997,shaver1999,tozzi2000,furlanetto2006a}. The global $21$-cm signal but is expected as a faint wideband feature in the sky brightness temperature at frequencies $1$--$200$~MHz \cite{pritchard2012} and no detection has been verified to date. Radio observations targeting this signal are being carried out by several experiments, most of which correspond to single-antenna, wideband, total-power radiometers \cite{bowman2018, philip2019, spinelli2021, deleraacedo2022, singh2022, monsalve2024, bull2025, mckay2025}. The antenna used by most experiments is a wide-beam monopole or dipole, which remains mechanically fixed above the soil and conducts observations as the sky transits overhead. Due to its small amplitude, detecting the $21$-cm signal requires hundreds of hours of integration and exquisite instrumental calibration. To avoid confusing the $21$-cm signal with any frequency-dependence, or chromaticity, of the beam, it is necessary to characterize the beam with high accuracy. Efforts are underway in the radio cosmology community to measure wide antenna beams with the required accuracy \cite{pober2012, neben2015, jacobs2017, bhopi2023, restrepo2023, herman2024, kuhn2025}. Until these efforts succeed, experiments will rely on electromagnetic (EM) simulations to determine the chromatic beam. The performance of wide-beam antennas operating above the soil is significantly influenced by the soil's EM properties \cite{smithrose1933, smithrose1935}. Therefore, the beam EM simulations must be computed including the geometry and EM properties of the soil, in addition to those of the instrument. The need for accurate beam simulations transforms the EM characterization of the soil at observation sites into a critical step for the calibration of ground-based global $21$-cm experiments \cite{spinelli2022, monsalve2024b, pattison2025}

In addition to radio cosmology, knowledge of the electrical properties of surface media is of keen interest to the geosciences community and adjacent fields. The electrical properties of surface media can be used to determine other physical characteristics of the media, therefore enabling a wide variety of applications in geology, hydrology, glaciology, archaeology, telecommunications, and civil engineering, among other areas \cite{lytle1974, IEEE_self_impedance_summary, reynolds2011}. Of particular relevance for our work is the characterization of cryospheric tundra soils in the Arctic. These soils include a shallow top layer that experiences seasonal thawing during summer months, known as the active layer, which is underlain by permafrost, defined as ground that remains frozen for two years or more \cite{harris1988,pollard2009, wilhelm2011}. Permafrost is recognized as a critical element of polar ecosystems: its sensitivity to climate change has important consequences for ecology and human infrastructure in the cryosphere, and increased thaw can generate positive feedback processes contributing to a warming climate  \cite{Meredith2019IPCC}. Knowledge of the physics and dynamics of permafrost, enabled by electrical characterization, is therefore a high priority for deepening our understanding of the cryosphere and the global climate \cite{kneisel2008, minsley2012}.

For a non-magnetic soil, its EM behavior can be characterized in terms of the complex relative permittivity $\epsilon=\epsilon_r-j\sigma(2\pi\epsilon_0 f)^{-1}$, where $\epsilon_r$ is the relative permittivity, $\sigma$ is the electrical conductivity, $\epsilon_0$ is the permittivity of vacuum, and $f$ is frequency. The conductivity and relative permittivity also intrinsically vary with frequency \cite{hoekstra1974, mironov2010}, and geophysically depend on the soil's mineral composition, structure, bulk density, temperature, water content, and organic content \cite{hipp1974, hallikainen1985, dobson1985, mccarter1997, francisca2003, mironov2015, shan2015}. Techniques traditionally used for soil electrical characterization include: (1) electrical resistivity tomography (ERT), which relies on measurements in the direct current regime \cite{herring2023}; (2) the capacitively-coupled resistivity (CCR) and (3) induced polarization (IP) techniques, both using measurements at kHz frequencies \cite{kuras2006, doetsch2015, mudler2022}; (4) ground-penetrating radar (GPR) and (5) time-domain reflectometry (TDR), using measurements at MHz--GHz frequencies \cite{topp1980, topp1985, boike1997, robinson2003, cataldo2008, leger2017}; and (6) resonant cavities and (7) coaxial line probes, used in laboratory characterization \cite{gorriti2005, mironov2010, bircher2016, zou2023}. 

One of the antenna parameters affected by the soil's electrical properties, in addition to the beam, is the impedance \cite{bhattacharyya1963, galejs1971, nicol1980, metwally1981, ridd1990, tokarsky2022, tokarsky2023}. This dependence can in turn be leveraged to characterize the soil at radio frequencies using the impedance of an antenna placed in the soil or above the surface \cite{IEEE_self_impedance_summary, Farzamian2020}. Impedance measurements are typically reported as a function of frequency, and therefore represent an attractive alternative to GPR and TDR, which conduct measurements in the time-domain. GPR measurements require a transmitting and a receiving antenna while impedance measurements only require a single antenna, similarly to the use of a single probe at the end of a transmission line in TDR. In early works, the impedance method was typically implemented using monopole, dipole, and loop antennas, for which analytical impedance models were available \cite{smith1974,wong1977,smith1985,smith1987,nicol1988}. In recent applications of the method, the impedance models have been generated through EM simulations instead of analytical expressions, opening up the technique to any antenna geometry \cite{wakita2000, lenlereriksen2004, berthelier2005, legall2006, cataldo2022, altamirano2025}. 

In this paper, we present estimates for the bulk electrical conductivity and relative permittivity at a site in the Canadian High Arctic used for sky observations by the Mapper of the IGM Spin Temperature (MIST) global $21$-cm experiment \cite{monsalve2024}. The soil electrical parameters are derived from impedance measurements of the antenna used by MIST for sky observations. The MIST instrument is a total-power radiometer with a single wideband dipole antenna mounted above the surface observing the sky at $25$--$125$~MHz. A key feature of the MIST approach is the choice to not include a metal ground plane beneath the antenna. Metal ground planes reduce the absolute sensitivity of antennas to the soil and for this reason are used by other global $21$-cm experiments \cite{bowman2018, philip2019, spinelli2021, deleraacedo2022, mckay2025}. However, metal ground planes could also introduce chromatic effects that complicate cosmological observations, which we prefer to avoid. More details about this experimental approach are provided in \citeA{monsalve2024} and \citeA{altamirano2025}. In the field, in addition to the sky spectrum, MIST conducts autonomous measurements of the antenna impedance, which are required for absolute radiometer calibration \cite{rogers2012, monsalve2017}. Here we use these impedance measurements across $25$--$125$~MHz in conjunction with impedance models from EM simulations for the critical purpose of estimating the electrical parameters of the soil, which are in turn required to compute EM simulations of the MIST antenna beam. The measurements were conducted on July $17$--$28$, 2022, and with a cadence of $111$~minutes. This cadence was established from requirements for radiometer calibration but here we explore if this cadence is also sufficiently high for tracking any time variability of the soil parameters. This work represents the first application of the antenna impedance technique for soil characterization at our observation site and with our full instrument. A proof-of-concept using a replica of our antenna and a basic measurement setup is described in \citeA{altamirano2025}. Due to the very small amplitude of the global $21$-cm signal, MIST must conduct sky observations and impedance measurements for long periods of time in an environment free from EM interference introduced by people or extra equipment in the vicinity. The technique demonstrated here is, therefore, particularly advantageous for MIST since it enables us to characterize the soil using autonomous measurements already included in our calibration program, thus avoiding the need for personnel or additional equipment devoted to this purpose. In this paper we describe the electrical characterization of our Arctic site and discuss how our best-fit soil model and parameters are expected to impact our cosmological results. We also argue that the autonomous and non-invasive nature of our approach makes it an attractive alternative to more traditional ground-based permafrost characterization techniques such as ERT and GPR.

\section{Study Site in the Canadian High Arctic}
\label{section_observing_site}
\begin{figure}
\centering
\includegraphics[width = 0.9\linewidth]{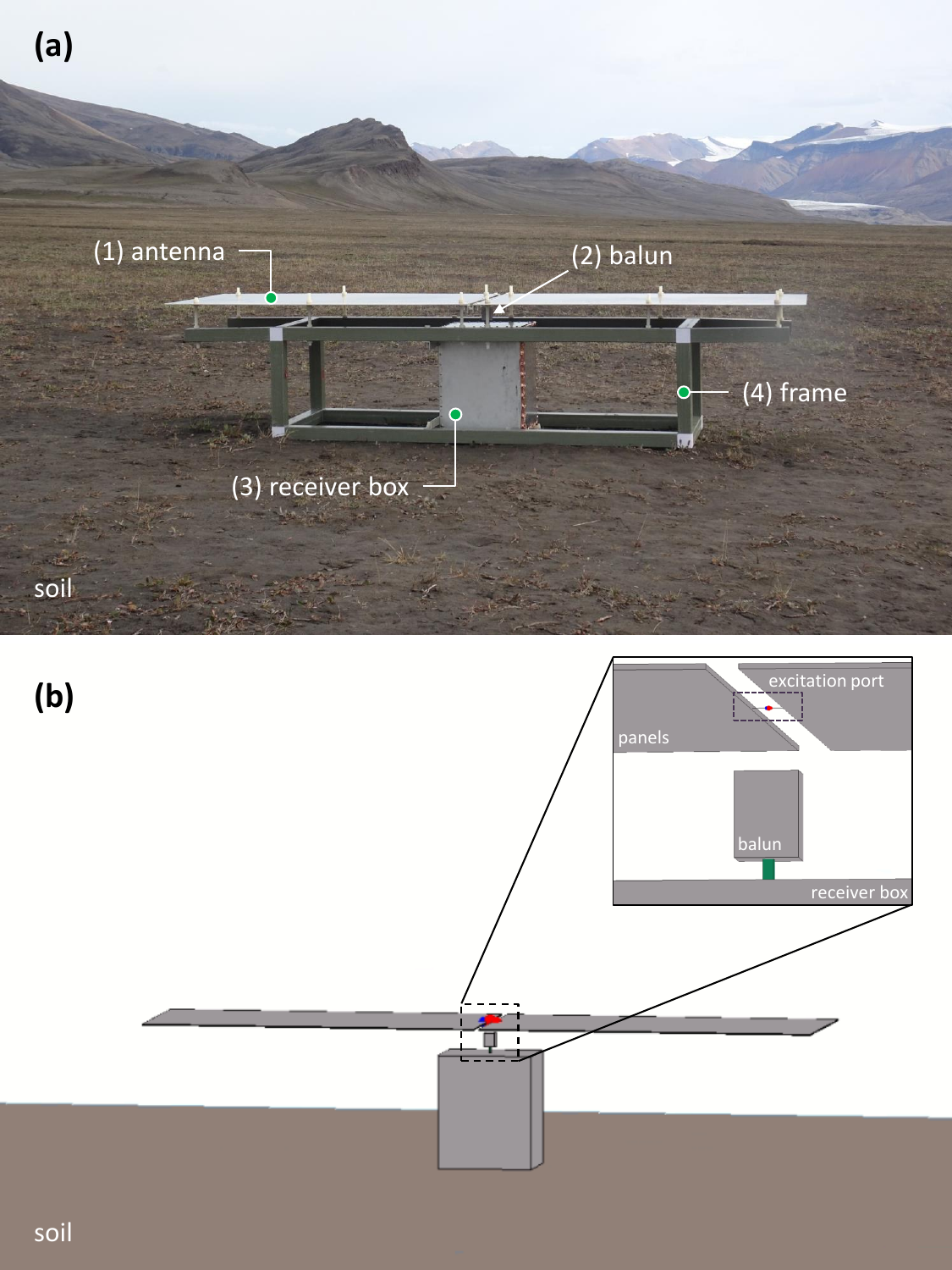}
\caption{(a) The MIST instrument at the study site in the Canadian High Arctic. The instrument consists of: (1) a horizontal blade dipole antenna formed by two aluminum panels of size $1.2$~m~$\times$~$60$~cm~$\times$~$3$~mm, separated horizontally by $2.15$~cm, and mounted $52$~cm above the soil; (2) a $5$-cm-long balun attached to the antenna excitation port at its input, and to the input of the receiver box at its output; (3) a $40.5$-cm-tall receiver box containing all the electronics (except for the balun) and four $12$-V batteries; (4) a support frame consisting of fiberglass and plastic parts. The antenna impedance measurements used in this analysis are calibrated at the antenna excitation port located between the two aluminum panels. (b) The simulated MIST instrument in Feko. The frame is not included in the simulations as it has a negligible contribution to the antenna impedance. The inset shows a zoomed-in view of the area around the excitation port. In Feko, the port is defined at the center of the two antenna panels and connected to them with two $1$-mm wires of length $1.075$~cm. These wires do not exist in the real instrument and their reactance is analytically removed from the simulated impedance before using the simulations to fit the measurements. The simulated soil extends to infinity in both horizontal directions and in depth, though this is not shown in the 3D view rendered by Feko.}
\label{figure_photo}
\end{figure}

The study site is located at Expedition Fjord on Umingmat Nunaat (Axel Heiberg Island), Nunavut, Canada, about $8$~km southwest of the McGill Arctic Research Station (MARS) \cite{pollard2009}. The site coordinates are $79.37980^{\circ}$~N, $90.99885^{\circ}$~W. The measurements were carried out on July $17$--$28$, 2022. Panel~(a) of Figure~\ref{figure_photo} shows MIST conducting measurements at the site. Figure~S1 in Supporting Information~S1 shows a map of the site. This site was selected because, due to its remoteness, it is not reached by most artificial radio emissions produced in urban centers, which would otherwise contaminate our sky observations \cite{dyson2021}.

The site is located on the Expedition River floodplain \cite{heldmann2005}, about $300$~m north of the point where the river, which flows westward downstream from the Thompson Glacier and White Glacier termini, meets Expedition Fjord. The geology of the site is characterized by alluvial-fan deposits \cite{mars_geology_Harrison2010}. The surface of the floodplain is relatively flat, homogeneous, and level. Ice-wedge polygons of varying width and length are less than $\approx 5$~cm in height \cite{haltigin2012}. The vegetation is relatively sparse, consisting primarily of short grasses and other small, low-lying plants, also under $\approx 5$~cm in height. To the best of our knowledge, no studies exist to date about the soil composition at our specific study site. However, \citeA{doran1993} did a stratisgraphic study of the lakebed at the center of Colour Lake, which is located at MARS. They report that the lakebed samples primarily consist of silt, followed by sand and clay, as well as $<20\%$ of organic matter. \citeA{rahman2019} found that silt was also the main soil constituent on and around a runway next to Colour Lake. Our site was affected by rain for several days until the day before our twelve-day measurement period, as well as during days $4$--$8$. For this reason, the soil was damp throughout the period. 

The active layer thickness (ALT) in permafrost environments is typically defined as the maximum depth of thaw at the end of the thaw season \cite{michaelides2019, clayton2021}. In the MARS region, the ALT is about $40$--$60$~cm and the permafrost is estimated to extend to a depth of $\approx400$--$500$~m, at which point the ground becomes unfrozen again \cite{pollard2009}.
The depth of thaw, or thaw depth, is defined as the minimum depth between the ground surface and frozen ground at any time during the thawing season \cite{harris1988}. For the measurements presented here, conducted in July during the active layer thaw period, part of the active layer of the floodplain was unfrozen and its thickness corresponds to the thaw depth. On July~$22$, 2022, we used a metal probe \cite{brown2000, boike2022} to measure the thaw depth at our site. Specifically, we measured at $28$ spots within a $30$-m radius from the exact location of the MIST instrument obtaining an average of $52.2$~cm and a sample standard deviation of $1.3$~cm. \citeA{rahman2019} did metal probe measurements of the thaw depth at three spots around MARS during late July and early August, 2018, obtaining values in the range $40.8$--$60.3$~cm.

\section{MIST Instrument and Measurements}
\label{section_impedance_measurements}
A detailed description of the MIST instrument is provided in \citeA{monsalve2024}. The instrument is shown in panel~(a) of Figure~\ref{figure_photo}. The instrument's antenna is a blade dipole formed by two aluminum panels of size $\approx 1.2$~m $\times$ $60$~cm $\times$~$3$~mm, which are mounted horizontally $\approx 52$~cm above the soil and separated by a horizontal distance of $2.15$~cm. A $5$-cm-tall balun, connected directly under the antenna, transforms the differential signal produced by the panels into a ground-referenced signal. This signal is then guided to the instrument's receiver which, along with the $12$-V batteries that power the system, is housed in a $40.5$-cm-tall aluminum box located under the balun. As indicated in Section~\ref{section_introduction}, MIST does not use a metal ground plane under the antenna. Figure~S2 in Supporting Information~S1 shows a schematic diagram of the instrument. Table~S1 in Supporting Information~S1 lists the exact dimensions of the instrument. 

The instrument carries out observations autonomously and continuously, and organizes them into 111-min blocks. In each block, the first $107$~s are devoted to impedance measurements. These measurements include a measurement of the antenna and measurements of internal open, short, and load (OSL) standards used for impedance calibration. The rest of the block is devoted to measurements of power spectral density (PSD) from the antenna, representing the observations of the sky, as well as from internal devices used for PSD calibration \cite{monsalve2024}. The impedance is measured with a Copper Mountain R60 (\url{https://coppermountaintech.com/vna/r60-1-port/}) one-port vector network analyzer incorporated into the MIST receiver. The measurements are done across the frequency range $1$--$125$~MHz, with a frequency resolution of $250$~kHz, an intermediate frequency bandwidth of $100$~Hz, and a power of $0$~dBm. In this paper, we only use impedance data from $25$--$125$~MHz because this is the range where the antenna is most efficient. Further, although the raw data resolution is $250$~kHz, the analysis is done with data points spaced by $1$~MHz to reduce the computation time of the numerical EM simulations used to generate the impedance models. A $1$-MHz resolution is sufficient to capture the features of the impedance across frequency. We estimate the soil parameters from the impedance calibrated at the excitation port of the antenna, which is located where the balun connects to the two aluminum panels. The first step in the calibration process involves calibrating the antenna measurements using the measurements of the internal OSL standards. The rest of the process is described in Text~S1 in Supporting Information~S1.

\begin{figure}
\noindent
\includegraphics[width = \linewidth]{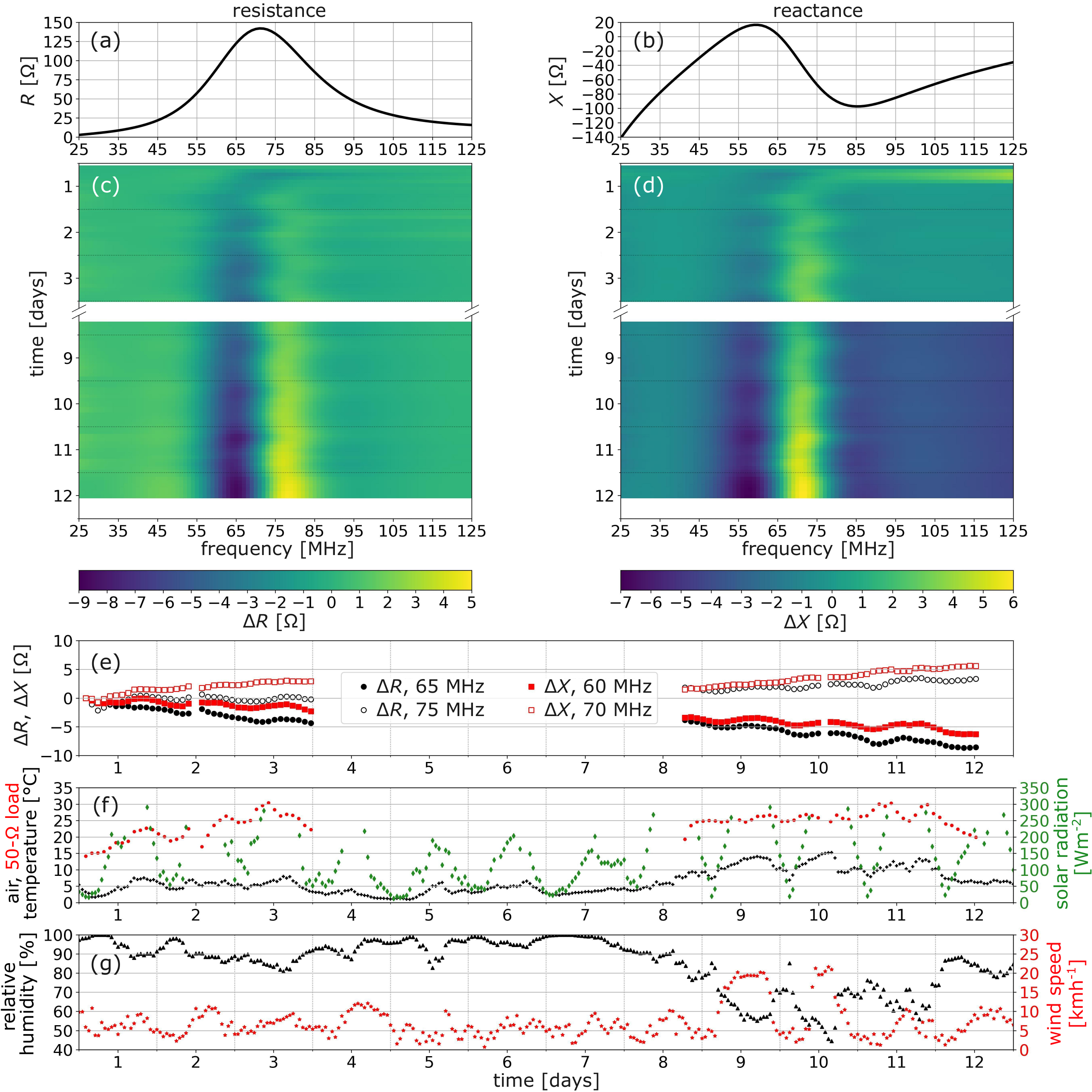}
\caption{Measurements of the MIST antenna impedance at the study site in the Canadian High Arctic. In the top row, panels~(a) and (b) show, respectively, the resistance and reactance of the first measurement. In the second row, panels~(c) and (d) show, respectively, the variations of the resistance and reactance across our twelve-day measurement period relative to the first measurement. On the $y$-axis of panels~(c) and (d), the ticks marking a given ``day" correspond to local noon, and the dashed horizontal lines to local midnight. No measurements were conducted between day~$4$ and the first three quarters of day~$8$ due to bad weather, and the final measurement ended halfway through day~$12$. Panel~(e) shows the variations of the resistance and reactance relative to the first measurement at two frequencies with large variations. Panel~(f) shows the temperature of the $50$-$\Omega$ load located inside the MIST receiver and used for impedance calibration, as well as air temperature and solar radiation measured by a weather station $\approx1$~km northwest of the study site. Panel~(g) shows the relative humidity and wind speed measured by the weather station.}    
\label{figure_impedance_measurements}
\end{figure}

The antenna impedance measurements calibrated at the excitation port are shown in panels~(a)--(e) of Figure~\ref{figure_impedance_measurements}. A total of $85$ measurements were conducted during the twelve-day observation period. We did not carry out measurements between day~$4$ and the afternoon of day~$8$ due to rain at the site. As panel~(a) shows, the antenna resistance (real part of the impedance) has a peak of $\approx140$~$\Omega$ at $\approx71$~MHz. Panel~(b) shows that the reactance (imaginary part of the impedance) is mostly capacitive, with a minimum of $\approx -140~\Omega$ at $25$~MHz. The reactance has a sinusoidal feature with a positive peak at $\approx 60$~MHz and a negative peak at $\approx 85$~MHz. Panels~(c) and (d) show that over the twelve-day period, the resistance and reactance evolve following a relatively monotonic trend. By day~$12$, the changes are up to $\approx 9$~$\Omega$ relative to the first measurement. In the time evolution, we also identify a daily modulation of amplitude $\lesssim 1~\Omega$. Measurements~$2$--$5$ of day~$1$ exhibit anomalous reactance above $\approx$~$75$~MHz, which we attribute to condensation on the antenna due to heavy fog affecting the site at that time. Panel~(e) shows the time evolution of the resistance and reactance at frequencies where the variations are most noticeable, and panels~(f) and (g) show temperature and other environmental variables providing context to the impedance measurements. Air temperature, solar radiation, relative humidity, and wind speed were recorded by an automatic weather station located $\approx$~$1$~km northwest of the study site, with the sensor measuring air temperature two meters above the ground surface in a solar radiation shield. Also shown is the physical temperature of the $50$-$\Omega$ load located inside the instrument's receiver box and used for internal OSL calibration. The temperature of the $50$-$\Omega$ load fluctuates because we do not control or stabilize the temperature in the receiver box, and is higher than the air temperature due to the heat produced by the receiver electronics. 

The variables plotted in panels~(f) and (g) of Figure~\ref{figure_impedance_measurements} do not show a monotonic increase or decrease in their time-evolution that could explain the monotonic patterns in the impedance. \citeA{ohmura1982} reports that evaporation is significant and the main water-losing process in the MARS region. We therefore suggest that the monotonic impedance patterns are driven by changes in the soil properties due to soil moisture variation caused by evaporation after the rain episodes that occurred before the start of our measurements and during days $4$--$8$. The sinusoidal fluctuations of the impedance on daily scales are seen to correlate most strongly with changes in solar radiation and air temperature. These environmental variables are not expected to significantly impact the soil properties on short time scales (see a more detailed discussion in Section~\ref{section_two_layer}), and therefore we regard as more likely the possibility that these patterns reflect changes in the instrument or measurement accuracy. The accuracy of the impedance calibration using the internal OSL depends primarily on the stability of the $50$-$\Omega$ load resistance, which in turn depends on temperature. We note that the $50$-$\Omega$ load temperature tracks the daily variations in the air temperature only partially. For this reason, if the daily impedance variations are of instrumental origin, the dominant cause is probably not related to changes in the internal OSL calibration. In principle, daily impedance variations could have occurred due to daily cycles of condensation onto the instrument, particularly the antenna panels, followed by evaporation. However, this explanation is unlikely considering that the daily pattern of impedance variation is very smooth and consistent across the study period while, as panels~(f) and (g) show, the solar radiation (which affects the temperature of the instrument) and, in particular, the air temperature and relative humidity (the factors that determine the dew point temperature) have very different values and patterns between chunks one and two.

\section{Analysis}
\label{section_analysis}
The soil parameter estimates and their uncertainties are obtained by fitting impedance models to each of the $85$ measurements using a $\chi^2$ minimization algorithm. The models are produced by interpolating between EM simulations of the impedance computed in advance on a regular grid of soil parameters. The simulations are computed with the Feko (\url{https://altair.com/feko}) software in the frequency range $25$--$125$~MHz at a resolution of $1$~MHz. The geometry of the Feko simulations is shown in panel~(b) of Figure~\ref{figure_photo}. The geometry includes the antenna panels, balun, receiver box, and soil. The soil extends to infinity in the horizontal directions and in depth, and is assumed to be flat in the horizontal directions. We consider two models for the soil, which differ in the number of vertical layers. The first model corresponds to a single-layer of homogeneous soil. The electrical parameters of the single-layer model are the bulk conductivity, $\sigma$, and relative permittivity, $\epsilon_{r}$. The second model is a stratified, two-layer soil. In the two-layer model, the parameters of the top layer are its conductivity, $\sigma_1$, relative permittivity, $\epsilon_{r1}$, and thickness, $t$. The parameters of the bottom layer are its conductivity, $\sigma_2$, and relative permittivity, $\epsilon_{r2}$. The two-layer model is motivated by the characteristics of soil in the High Arctic during the summer, which consists of an unfrozen soil at the top (the unfrozen part of the active layer) followed by frozen soil (the frozen part of the active layer and the permafrost underneath). 

The possibility of increasing the soil model complexity to make it more realistic has to be weighed against the higher computational cost expected during the analysis, as the number of simulations scales with the number of parameters (discussed in Text S2 in Supporting Information~S1). Increasing the number of vertical layers beyond two would in principle be justified by: (1) the presence at the study site of unfrozen ground below the permafrost at an estimated depth of $\approx400$--$500$~m \cite{pollard2009}, and (2) potential changes in the properties of the active layer as a function of depth due to changes in soil moisture or organic matter content \cite{chen2019, mironov2015, bakian_dogaheh_2022, bakian_dogaheh_2025}. Unfortunately, we anticipate that our experimental uncertainties will limit our sensitivity preventing us from going beyond a two-layer soil model. The surface of the soil at our site is not perfectly smooth as assumed in the simulations. However, the surface roughness is at the $\approx5$-cm level, which is significantly smaller than our measurement wavelengths ($\geq 2.4$~m). In addition, the gain of our antenna quickly approaches zero toward low elevation angles (i.e., the horizon). For these reasons, the surface roughness is expected to have a negligible effect and is not incorporated into our simulations. The properties of real soils are spatially heterogeneous and, therefore, our assumption of homogeneity within each soil model layer is in general inaccurate. However, in this work we test the hypothesis that, at our site, the electrical behavior of the soil layers can be well explained by bulk parameters and that the remaining effect from any spatial anisotropies is small. Finally, the electrical parameters of real soils in general vary as a function of frequency \cite{hoekstra1974, mironov2010}. However, here we assume that the dispersive behavior of the tundra soil is negligible and simulate the soil parameters as constant over $25$--$125$~MHz. To optimize the use of computational resources, we would incorporate into our soil models the refinements described above only if the fits indicate that it is mathematically justified. Details of the Feko simulations and the soil parameter grids are provided in Text~S2 and Table~S2 in Supporting Information~S1. The impedances simulated in Feko are shown in Figure~S3 in Supporting Information~S1. 

The interpolation between impedances simulated in Feko is done frequency by frequency, and separately for the resistance and reactance. The resistance and reactance are, at each frequency, smooth hyper-surfaces as a function of the soil parameters. Considering this smoothness, we interpolate the resistance and reactance at the required soil parameter combinations using cubic splines.

We include in the fits an additional free parameter to account for differences between the measured and modeled reactance that cannot be explained by the soil parameters or other known factors. We call this parameter the residual inductance $L$, and its effect is analytically incorporated in the fits by adding $X_L = 2 \pi f L$ to the impedance models, where $f$ is the frequency of the models. Considering $L$, the number of free parameters in the fits is three for the single-layer model and six for the two-layer model. The $\chi^2$ minimization algorithm used to fit the parameters accounts for the uncertainties in the impedance measurements and models. These uncertainties are described in Text~S3 and shown in Figure~S4 in Supporting Information~S1. More details about the fits are provided in Text~S4 in Supporting Information~S1.

\section{Results and Discussion}
\label{section_results}

\begin{figure}
\noindent
\includegraphics[width = \linewidth]{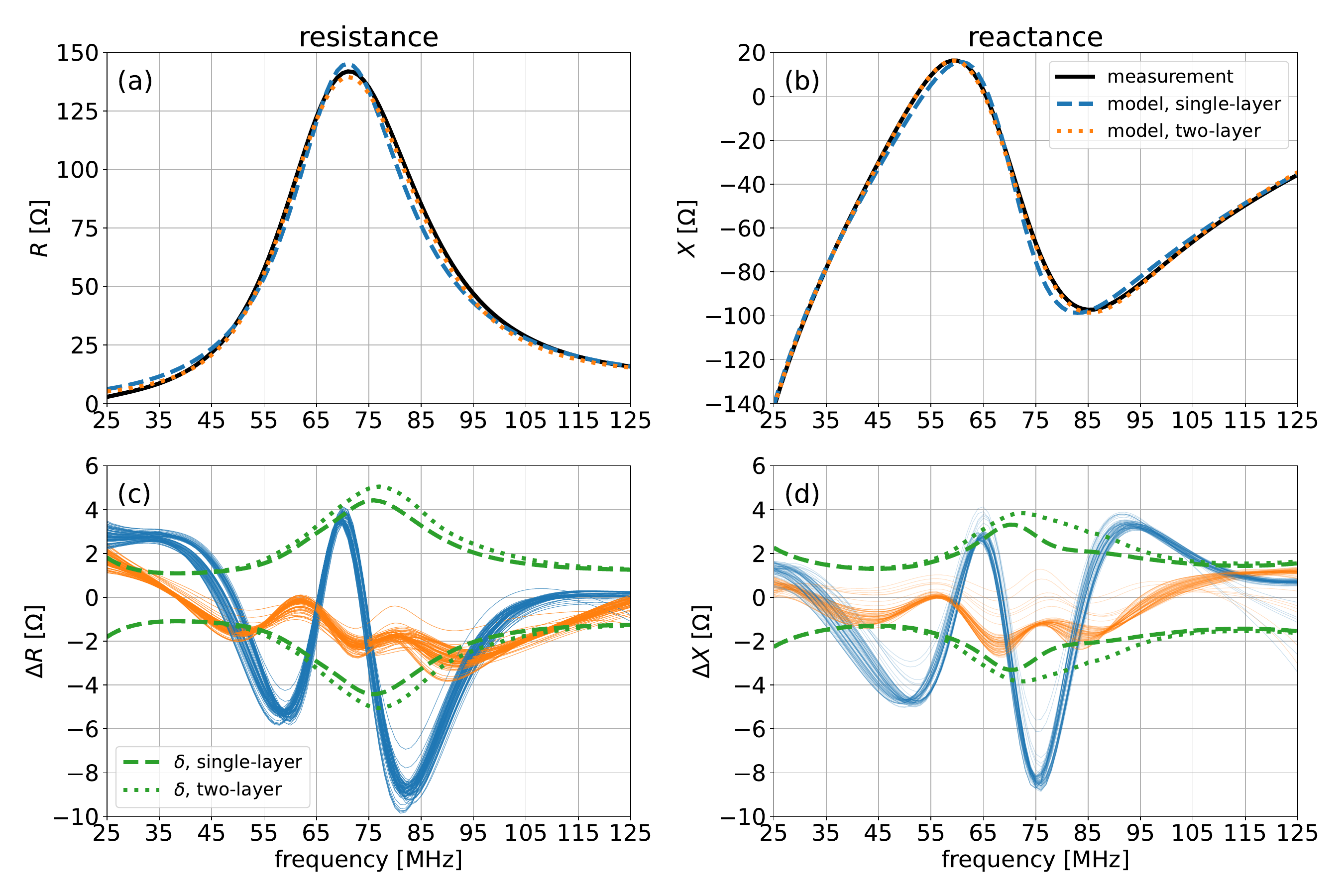}
\caption{Summary of the antenna impedance fits. Using the black line, panels~(a) and (b) show, respectively, the resistance and reactance from the first impedance measurement. The dashed blue and dotted orange lines show, respectively, the best-fit single- and two-layer impedance models. 
Panels~(c) and (d) show, for the resistance and reactance respectively, the fit residuals for all $85$~measurements. The blue (orange) lines show the residuals for the single-layer (two-layer) models. Panels~(c) and (d) also show the total uncertainty, $\delta$, accounted for in the fits. The green dashed (dotted) lines correspond to the total uncertainty for the single-layer (two-layer) fits.}
\label{figure_fits_and_residuals}
\end{figure}

Figure~\ref{figure_fits_and_residuals} presents a summary of the impedance fits. Panels~(a) and (b) show, as an example, the first measurement and best-fit single- and two-layer models. Panels~(c) and (d) show the residuals for the $85$ fits, as well as the $68\%$ total uncertainty used in the fits, which combines the measurement and model uncertainty.

\subsection{Single-layer Soil Model}
\label{section_single_layer}
In the single-layer case, the root mean square deviation (RMSD) of the fits, computed across frequency after concatenating the resistance and reactance residuals, is in the range $2.7$--$3.6$~$\Omega$. This RMSD is $\approx 1.9$--$2.6\%$ of the peak absolute value of the resistance and reactance, which is $\approx 140$~$\Omega$. The residuals show strong structure as a function of frequency and have a peak absolute value of $\approx 10$~$\Omega$. At some frequencies, the residuals are higher than the $68\%$ uncertainty by a factor of about three. The $\chi^2$ of the fits is in the range $688$--$1138$. In each fit, the number of data points considering resistance and reactance is $N_d=202$, and the number of free parameters is $N_p=3$. Therefore, the number of degrees of freedom is $\nu=N_d-N_p=199$. Values of the reduced chi-squared $\chi^2/\nu \geq (688/199=3.46)$ are high and indicate that the single-layer model does not fit the data well. To compare the single- and two-layer models we use the Bayesian information criterion (BIC), defined as $\chi^2+N_p\log N_d$ \cite{schwarz1978}. The BIC penalizes underfitting and overfitting, and a lower value indicates a better model. The BIC for the single-layer model is in the range $704$--$1154$.

\subsection{Improving the Fits with the Two-Layer Soil Model}
\label{section_two_layer}

The two-layer residuals are noticeably lower than in the single-layer case. They have a peak absolute value of $\approx 4$~$\Omega$ and remain within the $68\%$ uncertainty across most of the frequency range. The RMSD of the two-layer fits is $0.9$--$1.7$~$\Omega$, corresponding to $\approx 0.6$--$1.2\%$ of the peak resistance and reactance. The $\chi^2$ of the fits is $87$--$230$ and, since $N_d=202$ and $N_p=6$, the BIC is $119$--$262$, several times lower than for the single-layer model. On this basis, we choose the two-layer soil parameter estimates as our nominal results. The number of degrees of freedom is $\nu=196$ and the reduced chi-squared $\chi^2/\nu$ is in the range $0.44$--$1.17$. This range indicates that the two-layer model fits the data sufficiently well and that, considering the uncertainties in our impedance measurements and models, it is not mathematically justified to make our soil model more complex. Therefore, we refrain from adding more layers to the soil model, or spatial or frequency dependence to its electrical parameters.

\begin{figure}
\noindent
\includegraphics[width = \linewidth]{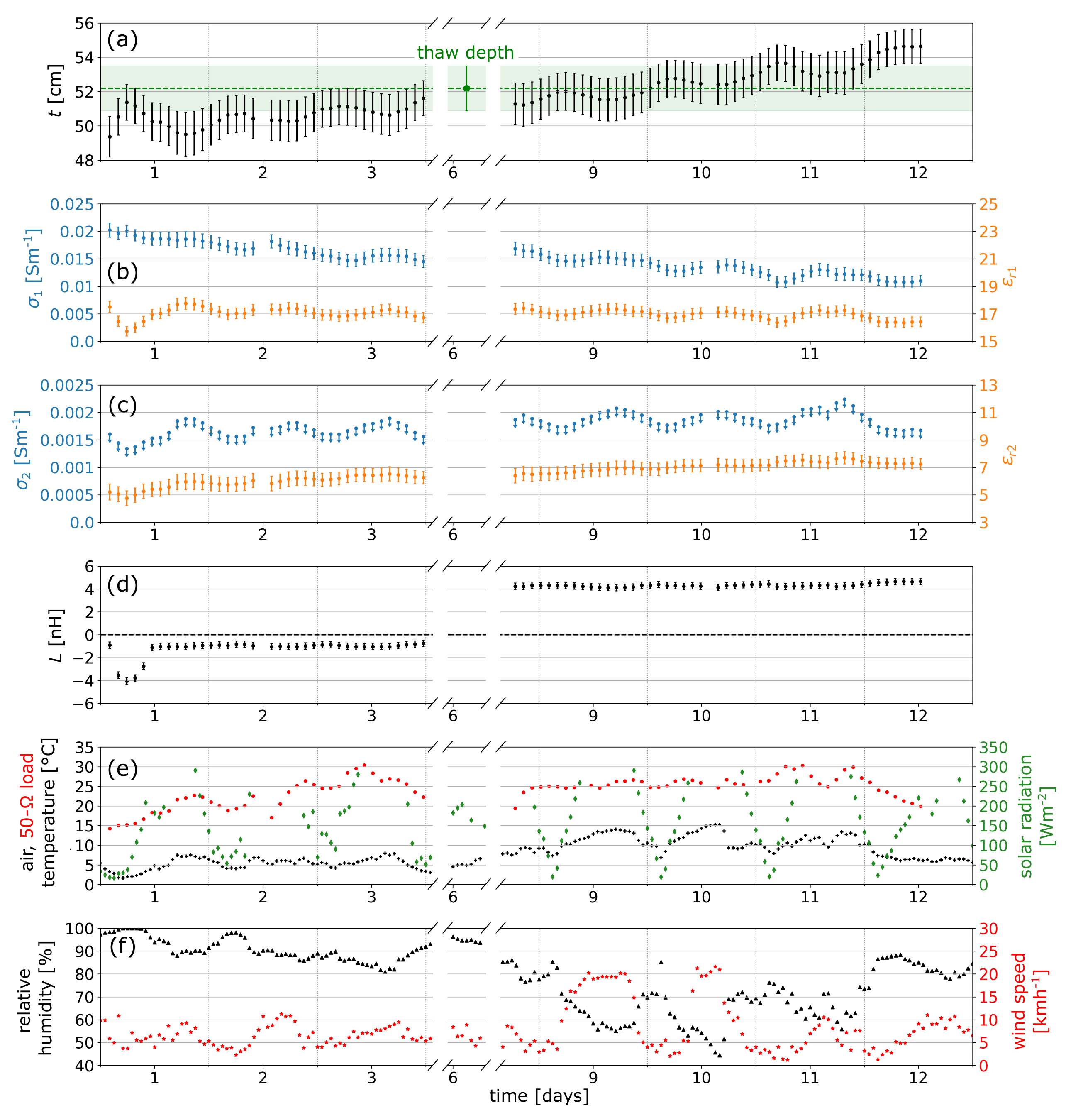}
\caption{Two-layer soil parameter estimates for the MIST site in the Canadian High Arctic. (a) Top-layer thickness, $t$. (b) Top-layer conductivity, $\sigma_1$, and relative permittivity, $\epsilon_{r1}$. (c) Bottom-layer conductivity, $\sigma_2$, and relative permittivity, $\epsilon_{r2}$. (d) Residual inductance, $L$. (e) Temperature of the internal $50$-$\Omega$ load, air temperature, and solar radiation. (f) Relative humidity and wind speed. In panel~(a), the green dot and error bars represent the average and sample standard deviation, respectively, of the $28$ thaw depth measurements done at the study site on July~22, 2022, with a metal probe. The green dashed line and band are used to extend, for reference, the green dot and error bars across the measurement period. In panel~(c), the blue downward-pointing arrows represent upper limits for $\sigma_2$ at the $68\%$ confidence level.}
\label{figure_results_two_layer}
\end{figure}

The soil parameter estimates obtained for the two-layer model are presented in Figure~\ref{figure_results_two_layer}. Panel~(a) shows that the top-layer thickness, $t$, spans $\approx 49.4$--$54.6$~cm in a predominantly ascending trend ($11\%$ increase). The $68\%$ uncertainty in these estimates is $\approx 1.1$~cm. For most of the study period, this top-layer thickness is consistent to within the uncertainties with the mechanical probe measurements of the thaw depth conducted on day~$6$, which yielded $52.2\pm1.3$~cm. In panel~(a), these thaw depth results are shown in green. Panel~(b) shows the top layer conductivity and relative permittivity. Both parameters follow a descending trend, with $\sigma_1$ decreasing from $\approx 0.020$~Sm$^{-1}$ to $0.011$~Sm$^{-1}$ ($45\%$ decrease), and $\epsilon_{r1}$ from $\approx 17.5$ to $16.4$ ($6\%$ decrease) with a dip to $\approx15.7$ for points $2$--$5$ of day~$1$ in chunk one. Panel~(c) shows the bottom layer results. The $\sigma_2$ estimates are consistent with zero to within their uncertainties and at $68\%$ confidence are constrained to $\lesssim 0.002$~Sm$^{-1}$. These $68\%$ upper limits are depicted in panel~(c) as downward-pointing blue arrows. The $\epsilon_{r2}$ estimates follow a predominantly ascending trend from $\approx 5.2$ to $7.2$ ($38\%$ increase). Panel~(d) shows the residual inductance. For chunk one, $L$ is typically $\approx-1$~nH. The exception is points $2$--$5$ of day~$1$, for which $L$ is as low as $\approx -4$~nH  seemingly due to the anomalous reactance in the corresponding measurements, as shown in Figure~\ref{figure_impedance_measurements}. For chunk two, $L$ is $\approx +4$~nH. Using as reference a wire of $1$-mm diameter, a self-inductance of $4$~nH corresponds to a wire length of $7.4$~mm. This length is small compared to the size of the instrument and reflects a high accuracy in the Feko simulations. The two-layer soil parameter estimates are summarized in Table~\ref{table_results}. The $68\%$ estimate uncertainties are very stable across measurements and, therefore, in the table we show their average. We also show the uncertainties as a percent precision relative to the best-fit value. With the exception of $\sigma_2$, for which we only determine upper limits, the precision of all our parameter estimates is $10\%$ or better.

\begin{table}
\caption{Summary of two-layer soil parameter estimates for the MIST site in the Canadian High Arctic}
\label{table_results}
\centering
\begin{tabular}{l      c c c c  c c c c c}
\hline
&  \multicolumn{4}{c}{Chunk One (July 17--19, 2022)} & \multicolumn{4}{c}{Chunk Two (July 24--28, 2022)} & \\
\cmidrule(lr){2-5}
\cmidrule(lr){6-9}
Parameter & Start & End & Min & Max & Start & End & Min & Max & 68\% uncertainty$^a$\\
\hline
$t$~[cm] & 49.4 & 51.6 & 49.4 & 51.6 & 51.3 & 54.6 & 51.2 & 54.6 & 1.1 (2\%) \\
$\sigma_1$~[Sm$^{-1}$] & 0.020 & 0.014 & 0.014 & 0.020 & 0.017 & 0.011 & 0.011 & 0.017 & 0.001 (5\%--9\%)\\
$\epsilon_{r1}$ & 17.5 & 16.7 & 15.7 & 17.8 & 17.4 & 16.4 & 16.3 & 17.4 & 0.4 (2\%--3\%)\\
$\sigma_2$~[Sm$^{-1}$] & & $\lesssim 0.002$ & (68\%) & & & $\lesssim 0.002$ & (68\%) & & --\\
$\epsilon_{r2}$ & 5.2 & 6.2 & 4.8 & 6.5 & 6.4 & 7.2 & 6.4 & 7.7 & 0.5 (6\%--10\%)\\
\hline
\multicolumn{10}{l}{$^{a}$ {Average $68\%$ uncertainty across both chunks and corresponding percent precision}.}
\end{tabular}
\end{table}

To quantify the sensitivity of the results to the residual inductance, we conduct a separate set of fits in which we estimate the soil parameters after fixing $L$ at zero. The results of this test are presented in Table~S3 in Supporting Information~S1. The changes in the soil parameters relative to the nominal values are quantified using the RMSD, which is computed across both data chunks. The table shows that when $L$ is fixed at zero the RMSD is consistent with the $68\%$ uncertainties of the nominal results for most of the parameters. The exception is $\sigma_1$, for which the RMSD is $0.002$~Sm$^{-1}$ while the uncertainty is $0.001$~Sm$^{-1}$. With nominal best-fit values in the range $\approx 0.011$--$0.020$~Sm$^{-1}$, an RMSD of $0.002$~Sm$^{-1}$ represents a bias of $\approx 10$\%--$18\%$ for this parameter. We avoid this bias in our nominal analysis by making $L$ a free parameter. The table also shows the results of an additional test done on the two-layer estimates. In this test, we quantify the sensitivity of the estimates to the value of the top-layer thickness. Specifically, we constrain $t$ to the range $50.9$--$53.5$~cm, corresponding to the average plus/minus one standard deviation of the thaw depth measurements. In this case, the RMSD obtained for all the soil parameters is smaller than their nominal $68\%$ uncertainties. The results from these two tests indicate that our nominal results are robust.

The time evolution of the two-layer soil parameter estimates can be reasonably well explained by the environmental conditions at the site if the two simulated layers are interpreted as the thawed and frozen layers of the real soil \cite{robinson1999, wu2013, thring2014, shan2015, liu2022}. The increase of $t$ during chunks one and two is consistent with the thawing of the frozen portion of the active layer. We note, however, that $t$ is approximately the same just before and just after the bad weather break of days $4$--$8$, which suggests a pause in the thawing. The average air temperature decreased from $\approx5^{\circ}$C in chunk one to $\approx3^{\circ}$C during the break (see panel~(f) of Figure~\ref{figure_impedance_measurements}). This decrease, coupled with the reduction of incoming shortwave solar radiation due to clouds (same figure and panel), would reduce the surface energy balance, and evaporation of soil moisture \cite{ohmura1982} would have a cooling effect on the active layer, preventing its thawing \cite{clayton2021}. The conductivity of fresh water can be as high as $\approx0.05$~Sm$^{-1}$ depending on its concentration of mineral salts \cite{borowiak2020,gapparov2023} and the relative permittivity of fresh water is about $80$ at ambient temperature \cite{buchner1999}. Our estimates for $\sigma_1$, and to a lesser extent $\epsilon_{r1}$, show a general decrease with time that is not directly correlated with any of the measured environmental variables. This decrease could be explained by a decrease of the water content in the thawed layer due to evaporation following the rain episodes at the site. This interpretation is supported by the increase of $\sigma_1$ and $\epsilon_{r1}$ at the start of chunk two relative to the end of chunk one following the rain of days $4$--$8$. Our estimates for $\epsilon_{r2}$ show a general increase, which could be explained by an increase in the water content of the frozen portion of the active layer due to thawing. The fact that $\epsilon_{r2}$ did not increase during days $4$--$8$ appears consistent with the interpretation that the frozen portion of the active layer did not thaw during that time.

Similarly to the impedance measurements, the soil parameter estimates show sinusoidal variations on daily scales. To highlight these variations, we remove best-fit slopes from the estimates from chunks one and two, and show the residuals in Figure~S5 in Supporting Information~S1. When comparing panels~(a)--(d) with panel~(f), we note that the sinusoidal variations in the soil parameter estimates strongly follow the variations in solar radiation and air temperature. However, we do not expect that the real soil, especially the bottom layer, can respond to changes in solar radiation and air temperature on such short time scales. Further, if the variations in the soil parameter estimates were due to changes in the soil moisture we would expect $t$ to increase with solar radiation and air temperature. However, in the figure we see that $t$ is negatively correlated with these two quantities. These reasons suggest that the daily variations of the soil parameter eatimates are of instrumental origin. The specific origin, however, remains unknown and is most likely not dominated by fluctuations in the internal OSL calibration, as mentioned in Section~\ref{section_impedance_measurements}. The slope fit residuals for $t$ are typically within $\pm 0.5$~cm; for $\sigma_1$ within $\pm 0.001$~Sm$^{-1}$; for $\epsilon_{r1}$ within $\pm 0.5$; and for $\epsilon_{r2}$ within $\pm 0.25$. These residuals, except for points $2$--$5$ of day~$1$ and a few other points, are consistent with zero to within the $68\%$ uncertainties. Since the soil parameter fits minimize $\chi^2$, the uncertainties of the estimates returned by the fits depend on the uncertainties assigned to the measurements and models (Section~\ref{section_analysis}). The fact that the uncertainties of the estimates can account for the daily variations, whose specific origin is unknown, indicates that our uncertainty quantification is realistic.

\subsection{Consistency with Previous Estimates for Arctic Soils}
Verifying that the antenna impedance technique yields valid soil parameter results is critical before relying on this technique for $21$-cm cosmology. Our estimates for $\sigma_1$ and $\sigma_2$, of $>0.01$~Sm$^{-1}$ and $\lesssim0.002$~Sm$^{-1}$, respectively, are consistent with several laboratory measurements of unfrozen and frozen soil \cite{wu2013, shan2015, liu2022}. Our results for $\sigma_1$, $\sigma_2$, and $t$, are further consistent with several field measurements in the Arctic using ERT, CCR, IP, and GPR. Although ERT, CCR, and IP measurements are done at frequencies $\lesssim$~kHz, they still represent useful references for our results at $25$--$125$~MHz. These field measurements have consistently reported conductivities $\gtrsim 0.002$~Sm$^{-1}$ and $\lesssim0.002$~Sm$^{-1}$ for the thawed and frozen layers, respectively, which align with our results \cite{fortier1994, angelopoulos2013, dafflon2016, leger2017, tavakoli2021, rossi2022, uhlemann2023}.

Our results for $\epsilon_{r1}$ and $\epsilon_{r2}$ are close to the findings of \citeA{thomson2013}, who carried out GPR measurements at $450$~MHz in the summer at a floodplain near Thomas Lee Inlet, Devon Island, Nunavut, Canada ($75.358^{\circ}$~N, $88.681^{\circ}$~W). From common-midpoint surveys conducted along two transects, they report relative permittivities of $12$--$16$ and $2.8$--$4.7$ for the thawed and frozen layers, respectively. \citeA{wong1977} conducted the study that so far aligns the most with our work. They characterized the active layer at a site near Tuktoyaktuk, Northwest Territories, Canada ($69.433^{\circ}$~N, $133.033^{\circ}$~W), using impedance measurements of a monopole antenna inserted into the soil at frequencies up to $111$~MHz. In the summer, for the thawed active layer they determined a thaw depth of $\approx50$~cm, a conductivity of $0.012$--$0.12$~Sm$^{-1}$ and a relative permittivity of $26$--$64$. In the winter, for the frozen active layer the conductivity was constrained to $<3\times10^{-4}$~Sm$^{-1}$ and the relative permittivity was estimated at $4.2$--$7.0$. Their results are consistent with our estimates across a similar frequency range.

In addition to our own metal probe measurements on June~$22$, 2022, our estimates for $t$ are consistent with \citeA{rahman2019} who report a thaw depth at MARS in the range $40.8$--$60.3$~cm for late July and early August, 2018, during a similar seasonal period. The daily rate of increase in our top-layer thickness estimate is $\approx 0.5$~cm~d$^{-1}$. As a metric for the dynamics of the active layer, the thawed layer thickness daily increase rate is highly sensitive to local-scale effects (such as topography, weather, snow cover, vegetation, soil composition, soil moisture, and soil organic content), latitude, seasonal period, and intra-annual variability, and caution is therefore warranted when comparing results from different studies \cite{hinkel2003,yi2018,dobinski2020,li2022,liu2023}. Nevertheless, our results are broadly consistent with reports for other Arctic and sub-Arctic sites \cite{tran2018, kim2021, muir2025}. Future work could include comparison of our results with existing thaw depth and ALT modeling approaches, including the numerical heat transfer model introduced by \citeA{ling2004} and GIPL2-MPI \cite{gipl2}.

\subsection{Implications for $21$-cm Cosmology}
In \citeA{monsalve2024b} we found through simulations that the global $21$-cm signal extraction accuracy strongly depends on the relative parameter values between the soil model layers. For example, if the conductivity of the top layer is low and lower than in the bottom layer, radio signals would penetrate deeper into the soil and the bottom layer would produce stronger reflections back toward the antenna, complicating the extraction. For this reason, a lower bottom-layer conductivity would be preferred. The situation is reversed for the permittivity, where a higher bottom-layer permittivity would be preferred. The conditions at our study site are favorable considering the conductivity, which is lower in the bottom layer, and disfavorable considering the permittivity, which is also lower in the bottom layer. According to \citeA{monsalve2024b}, for these soil characteristics and our reported parameter values, the accuracy of the estimates must be of at least $10\%$, and better than $5\%$ for $\epsilon_{r1}$ and $t$. As Table~\ref{table_results} shows, we report a percent precision of $2\%$--$3\%$ for $\epsilon_{r1}$ and $t$, and $5\%$--$10\%$ for $\sigma_1$ and $\epsilon_{r2}$, thus satisfying the requirements for these parameters. For $\sigma_2$ here we only determine upper limits and, therefore, improving the constraints on this parameter becomes a priority toward the future. Our results nonetheless indicate that using the antenna impedance technique is a promising strategy to characterize our observation site and achieve competitive cosmological results. 

In the antenna impedance technique, increasing the accuracy and precision of the soil parameter estimates requires decreasing the uncertainty of the impedance measurements and models. In the future we will decrease the measurement uncertainty by developing temperature-dependent models for the balun and the impedance measurement subsystem inside the MIST receiver. We will decrease the model uncertainty through a denser sampling of the parameter space with precomputed EM simulations, as well as with more accurate measurements of the physical dimensions and orientation of the instrument at the site. 

In a twelve-day period our soil parameter estimates vary by more than a few percent. Larger variations would be expected for longer sky observation campaings due to the geophysical and environmental context at the site. Considering our accuracy requirements, the soil parameter values cannot be assumed to be stable during these periods. Figure~\ref{figure_results_two_layer} shows that, except for $\sigma_2$, the variation of the estimates between consecutive samples at $111$-min cadence is typically smaller (better) than the accuracy requirements in \citeA{monsalve2024b}. The $111$-min cadence thus appears to be sufficiently high for our application.

\section{Conclusion}
The electrical characterization of the soil is critical for ground-based single-antenna radio experiments trying to detect the global $21$-cm signal from the early Universe. Here we presented the first electrical characterization of the observation site used by the MIST experiment in the Canadian High Arctic. This characterization was done using measurements of the MIST antenna impedance at $25$--$125$ MHz carried out on July $17$--$28$, 2022, with a cadence of $111$~minutes. Relying on the antenna impedance for soil characterization is very advantageous for MIST because, due to its use for absolute radiometer calibration, the impedance is already included in the calibration program of the experiment. Using a $\chi^2$ minimization algorithm, the impedance measurements were fitted with models produced by interpolating between precomputed EM simulations of the MIST instrument and soil. We initially modeled the soil as a homogeneous medium but obtained impedance fits with high $\chi^2$. For this reason we then explored a two-layer soil model, where the bottom layer extends to infinite depth. The two-layer model produced lower $\chi^2$ and fit residuals consistent with our experimental uncertainties. These metrics, in addition to a lower BIC, made us choose the two-layer estimates as our nominal results. Our two-layer estimates are consistent with expectations for the Arctic tundra soil in the summer, which includes an unfrozen top layer (unfrozen part of active layer) and frozen soil beneath (frozen part of active layer followed by permafrost). Our estimates are further consistent with results obtained in the Arctic with techniques such as ERT, CCR, IP, and GPR. For MIST, this consistency represents valuable verification of the antenna impedance technique, which is not as widely-used for soil characterization. Except for the bottom-layer conductivity, for which we determined an upper limit, our soil parameter estimates were obtained with a percent precision of $10\%$ or better. This precision is close to the requirements for the accurate extraction of a typical global $21$-cm signal model. During our twelve-day study period, some parameters varied by more than a few percent. Considering our accuracy requirements, it is important to track the soil parameter changes through high cadence measurements. Our $111$~minute cadence proved sufficiently high for this purpose. In the near term we will use the soil parameter estimates reported here to calibrate the sky observations conducted by MIST during the same time period and attempt the extraction of the $21$-cm signal. Looking forward, we will work toward reducing our experimental uncertainties in order to improve our soil modeling and the calibration of future sky observations.

We finally highlight the potential of this technique to broadly contribute to the geosciences, noting its autonomous, non-invasive nature and spatial scalability, especially with stripped-down instruments that do not observe the sky but are only dedicated to impedance measurements. Such stripped-down instruments would operate more efficiently than traditional implementations of ERT and GPR, enabling this technique to provide long-term, high-cadence monitoring of soil with minimal supervision. We here show that the antenna impedance technique is capable of characterizing with precision (for most of the parameters) the thawed and frozen layers of the tundra soil in the Arctic summer. We thus anticipate that this technique can be at least as effective for the characterization of other layered soils \cite{kettridge2008, uhlemann2016}, especially where the electrical parameters are very different between the layers. Larger parameter differences would imprint stronger features on the impedance as a function of frequency, which in turn would increase our leverage when estimating the parameters. One potential application is the characterization and monitoring of frozen lakes and sea ice, taking advantage of the very different electrical properties between the surface ice and water below \cite{liu2014,balkhanov2018}. This technique could be used in isolation as the core of long-term monitoring programs but also as part of broader efforts, such as to complement remote sensing measurements \cite{barnes2003,johansson2017,babaeian2019,duncan2020}.

\section{Conflict of Interest}
The authors declare no conflicts of interest relevant to this study.

\section{Open Research}
The MIST impedance measurements, Feko simulations, and Python codes used to estimate and plot the soil properties are made publicly available in \citeA{hendricksen_2025_17117364}.

\section{Inclusion in Global Research Policy}
The field activities at the McGill Arctic Research Station (MARS) are undertaken with the permission of the Government of Nunavut through a research license (NPC File \#149098) and a land use permit to McGill University.

\acknowledgments
We are very grateful to the editors and reviewers for helping us to significantly improve this paper. We are grateful to Matt Dobbs for lending us equipment from his Cosmology Instrumentation Laboratory at McGill University. We are also grateful to Jo\"{e}lle Begin, Cherie Day, Eamon Egan, Larry Herman, Marc-Olivier Lalonde, and Tristan M\'enard for their support with operations in the Arctic. We are extremely grateful to the MARS directors and researchers for their invaluable advice and field support. We also acknowledge the Polar Continental Shelf Program for providing funding and logistical support for our research program, and we extend our sincere gratitude to the Resolute staff for their generous assistance and bottomless cookie jars. We acknowledge support from ANID Fondo 2018 QUIMAL/180003, Fondo 2020 ALMA/ASTRO20-0075, Fondo 2021 QUIMAL/ASTRO21-0053, and Fondo 2022 ALMA/31220012. We acknowledge support from ANID FONDECYT Iniciaci\'on 11221231 and ANID FONDECYT Regular 1251819. We acknowledge support from Universidad Cat\'olica de la Sant\'isima Concepci\'on Fondo UCSC BIP-106. We acknowledge the support of the Natural Sciences and Engineering Research Council of Canada (NSERC), RGPIN-2019-04506, RGPNS 534549-19. We acknowledge the support of the Canadian Space Agency (CSA) [21FAMCGB15]. We acknowledge support from the National Geographic Society (NGS-94983T-22). This research was undertaken, in part, thanks to funding from the Canada 150 Research Chairs Program. This research was enabled in part by support provided by SciNet and the Digital Research Alliance of Canada.

\clearpage

\bibliography{references}
% \printbibliography

\newpage

\title{Supporting Information for ``Estimating Soil Electrical Parameters in the Canadian High Arctic from Impedance Measurements of the MIST Antenna Above the Surface"}
%
% e.g., \title{Supporting Information for "Terrestrial ring current:
% Origin, formation, and decay $\alpha\beta\Gamma\Delta$"}
%
%DOI: 10.1002/%insert paper number here%

%% ------------------------------------------------------------------------ %%
%
%  AUTHORS AND AFFILIATIONS
%
%% ------------------------------------------------------------------------ %%

\authors{\
I. Hendricksen\affil{1}, \
R. A. Monsalve\affil{2,3,4}, \
V. Bidula\affil{1}, \
C. Altamirano\affil{4}, \
R. Bustos\affil{4}, \
C. H. Bye\affil{5}, \
H. C. Chiang\affil{1,6}, \
X. Guo\affil{2}, \
F. McGee\affil{1}, \
F. P. Mena\affil{7}, \
L. Nasu-Yu\affil{1}, \
C. Omelon\affil{8}, \
S. E. Restrepo\affil{4,9}, \
J. L. Sievers\affil{1,10}, \
L. Thomson\affil{8}, \
N. Thyagarajan\affil{11}}

\affiliation{1}{Department of Physics and Trottier Space Institute, McGill University, Montr\'eal, QC H3A 2T8, Canada}
\affiliation{2}{Space Sciences Laboratory, University of California, Berkeley, CA 94720, USA}
\affiliation{3}{School of Earth and Space Exploration, Arizona State University, Tempe, AZ 85287, USA}
\affiliation{4}{Departamento de Ingenier\'ia El\'ectrica, Universidad Cat\'olica de la Sant\'isima Concepci\'on, Alonso de Ribera 2850, Concepci\'on, Chile}
\affiliation{5}{Department of Astronomy, University of California, Berkeley, CA 94720, USA}
\affiliation{6}{School of Chemistry and Physics, University of KwaZulu-Natal, Durban, South Africa}
\affiliation{7}{National Radio Astronomy Observatory, Charlottesville, VA 22903, USA}
\affiliation{8}{Department of Geography, Queen's University, Kingston, ON K7L 3N6, Canada}
\affiliation{9}{Centro de Energ\'ia, Universidad Cat\'olica de la Sant\'isima Concepci\'on, Alonso de Ribera 2850, Concepci\'on, Chile}
\affiliation{10}{School of Mathematics, Statistics, \& Computer Science, University of KwaZulu-Natal, Durban, South Africa}
\affiliation{11}{Commonwealth Scientific and Industrial Research Organisation (CSIRO), Space \& Astronomy, P. O. Box 1130, Bentley, WA 6102, Australia}

% \begin{article}

\noindent\textbf{Contents of this file}
\begin{enumerate}
\item Text S1 to S4
\item Figures S1 to S5
\item Tables S1 to S3
\end{enumerate}

\noindent\textbf{Introduction}

\begin{figure}[h]
% \setfigurenum{S1} 
\centering
\includegraphics[width = 
\linewidth]{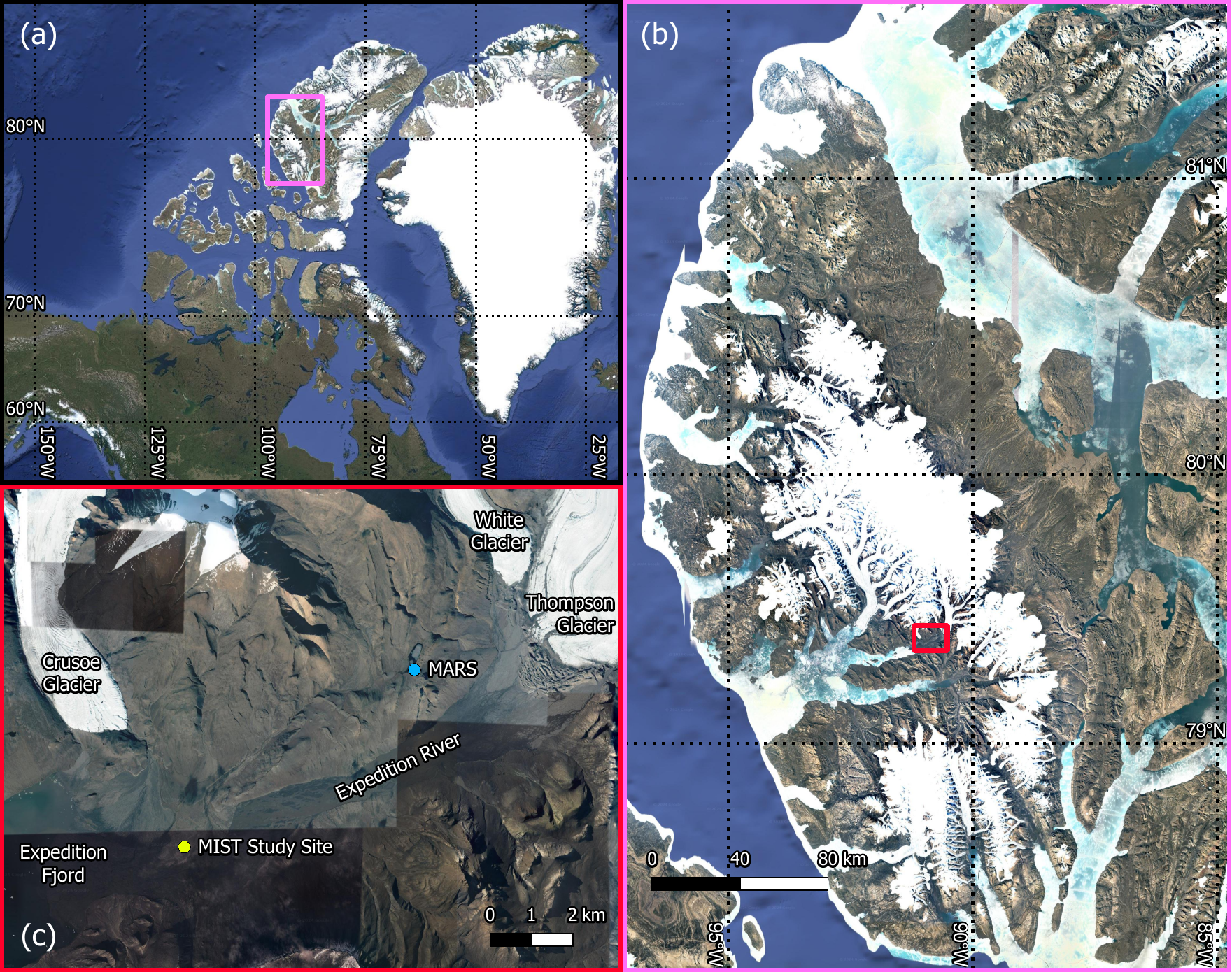}
\caption{Map of the MIST study site ($79.37980^{\circ}$~N, $90.99885^{\circ}$~W) in Expedition Fjord, Umingmat Nunaat (Axel Heiberg Island), Nunavut, Canada. This site is accessible from the McGill Arctic Research Station (MARS). (a) Satellite image of the Canadian High Arctic and Greenland, with Umingmat Nunaat (Axel Heiberg Island) identified with a pink box. (b) Umingmat Nunaat (Axel Heiberg Island), with red box identifying the MARS region. (c) The MARS region, with the MIST study site and MARS identified as yellow and blue dots, respectively. Maps created using Google Satellite data in the QGIS software.}
\label{figure_map}
\end{figure}

\begin{figure}
\centering
\includegraphics[width = 0.75\linewidth]{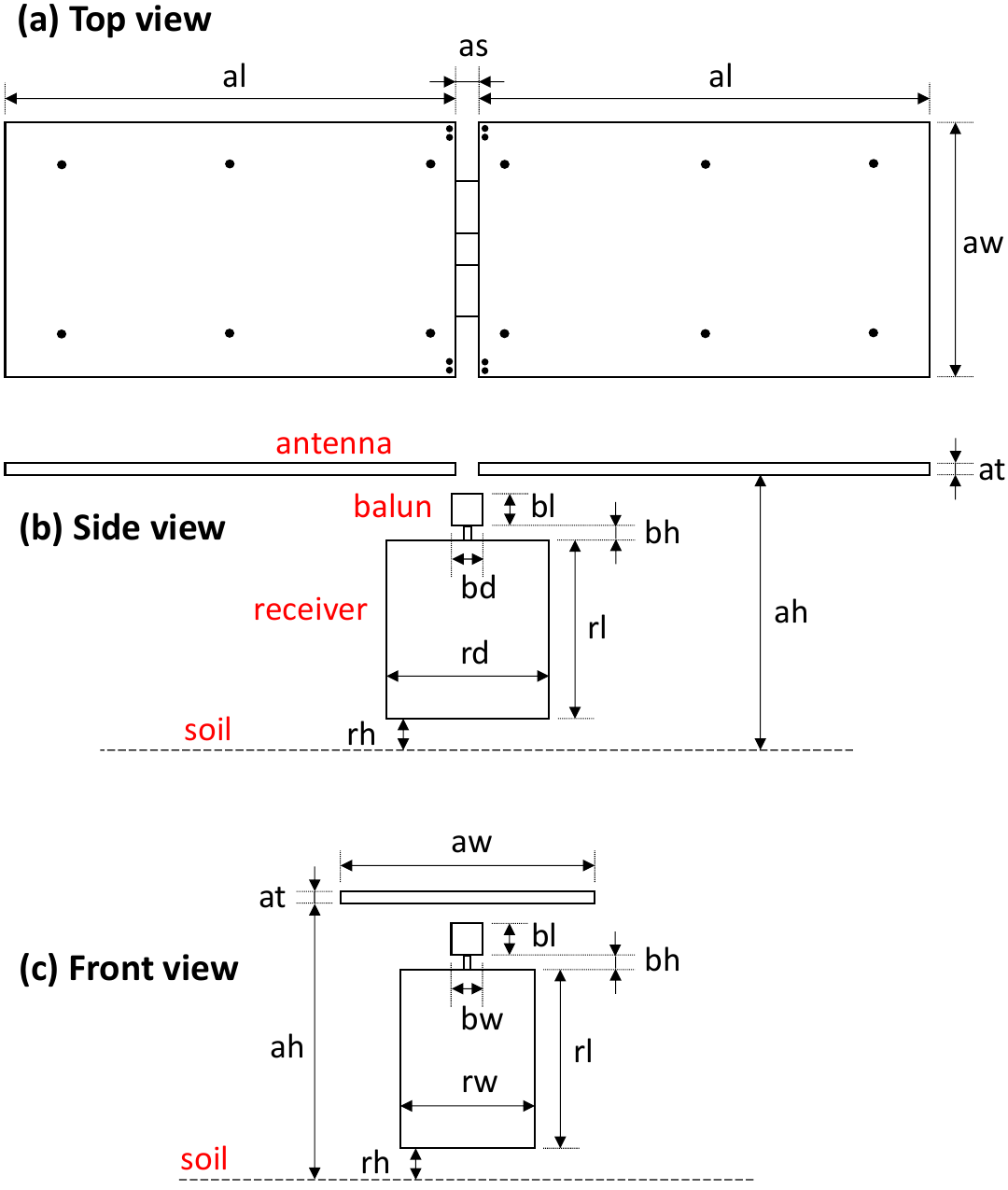}
\caption{Diagram of the MIST instrument with parameters indicating the physical dimensions. The parameter values are listed in Table~\ref{table_instrument_dimensions}. This figure is reproduced for reference from \citeA{monsalve2024}.}
\label{figure_instrument_dimensions}
\end{figure}

\begin{figure}
\noindent
\includegraphics[width = \linewidth]{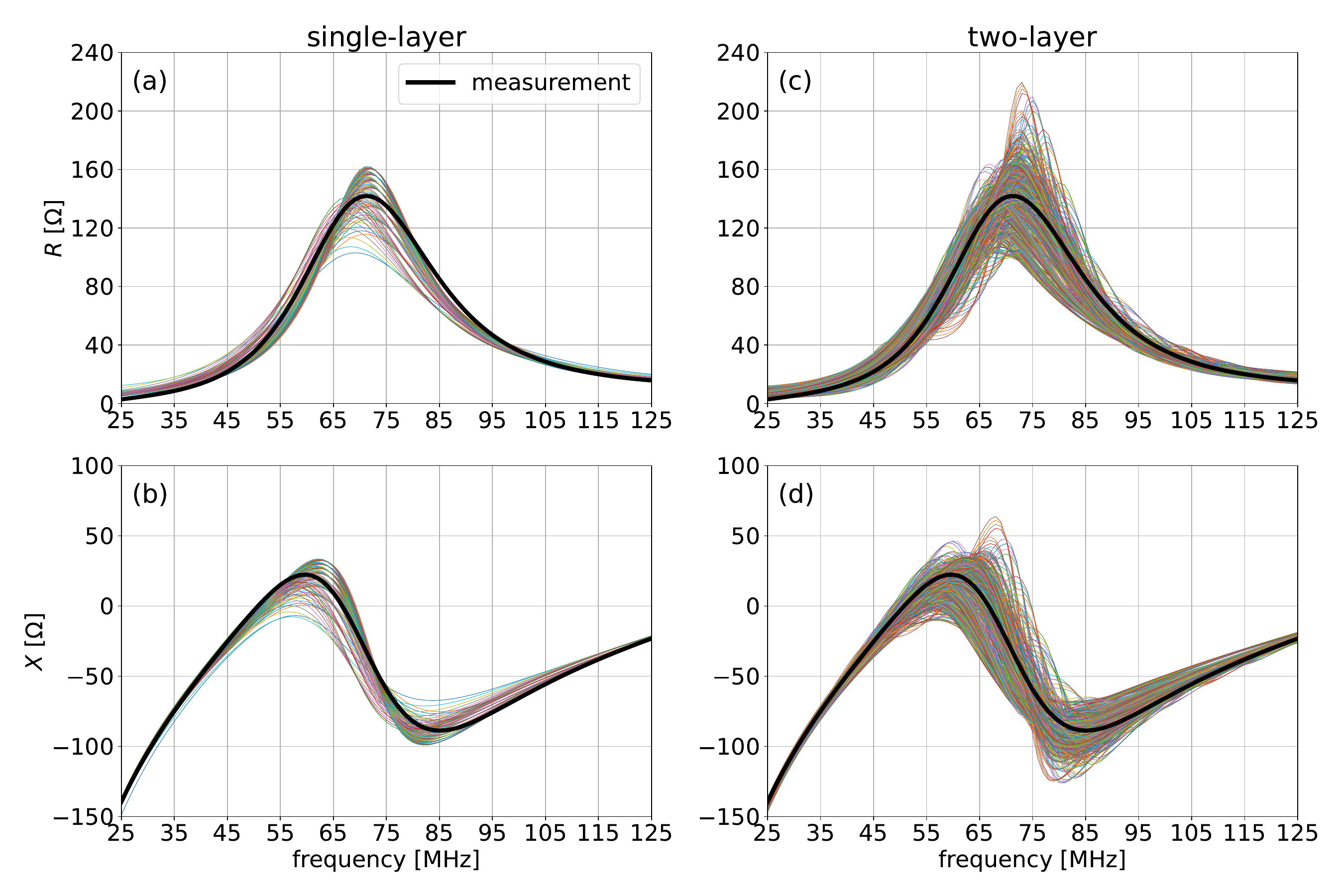}
\caption{The colored lines represent all the impedances simulated with Feko on the regular soil parameter grids defined by the information listed in Table~\ref{table_feko_simulations}. There are $81$~total single-layer simulations and $1024$~two-layer simulations. Panels~(a) and (b) represent, respectively, the resistance and reactance of the single-layer simulations. Panels~(c) and (d) represent, respectively, the resistance and reactance of the two-layer simulations. The thick black lines represent the first impedance measurement, which is shown for reference.}
\label{figure_feko_simulations}
\end{figure}

\begin{figure}
\noindent
\includegraphics[width = \linewidth]{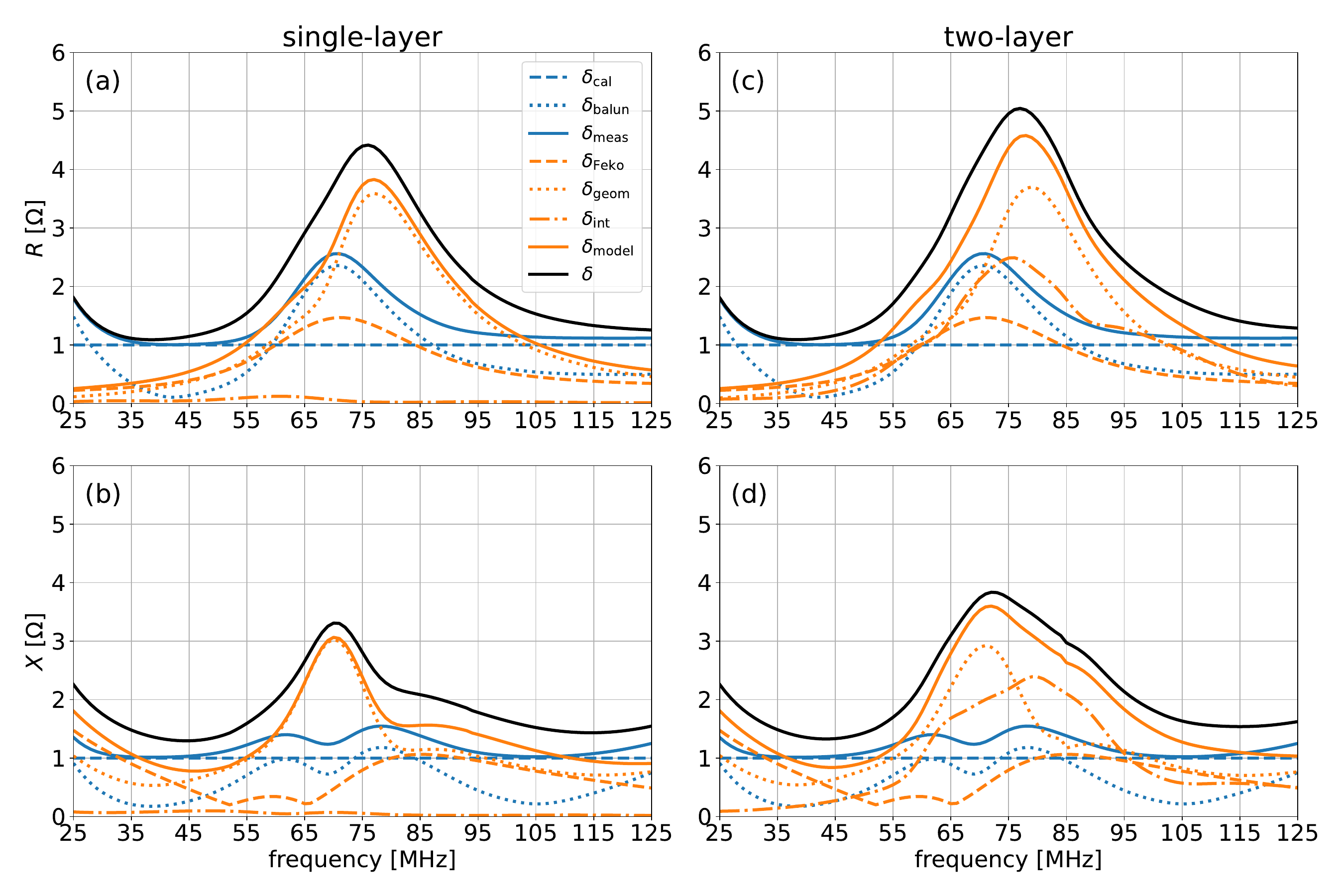}
\caption{Uncertainties reported at the $68\%$ confidence level for the single-layer and two-layer models. (a) Single-layer resistance uncertainties. (b) Single-layer reactance uncertainties. (c) Two-layer resistance uncertainties. (d) Two-layer reactance uncertainties. The description of the uncertainties is provided in Text~S3. The total uncertainty, $\delta$ (solid black), is computed as the quadrature sum of the impedance measurement uncertainty, $\delta_{\textrm{meas}}$ (solid blue), and the impedance model uncertainty, $\delta_{\textrm{model}}$ (solid orange). The measurement uncertainty is computed as the quadrature sum of the receiver calibration uncertainty, $\delta_{\textrm{cal}}$ (dashed blue), and the balun S-parameter uncertainty, $\delta_{\textrm{balun}}$ (dotted blue). The model uncertainty is computed as the quadrature sum of the Feko accuracy uncertainty, $\delta_{\textrm{Feko}}$ (dashed orange), the Feko geometry uncertainty, $\delta_{\textrm{geom}}$ (dotted orange), and the model interpolation uncertainty, $\delta_{\textrm{int}}$ (dash-dotted orange).}
\label{figure_total_uncertainty}
\end{figure}

\begin{figure}
\noindent
\centering
\includegraphics[width = 0.93\linewidth]{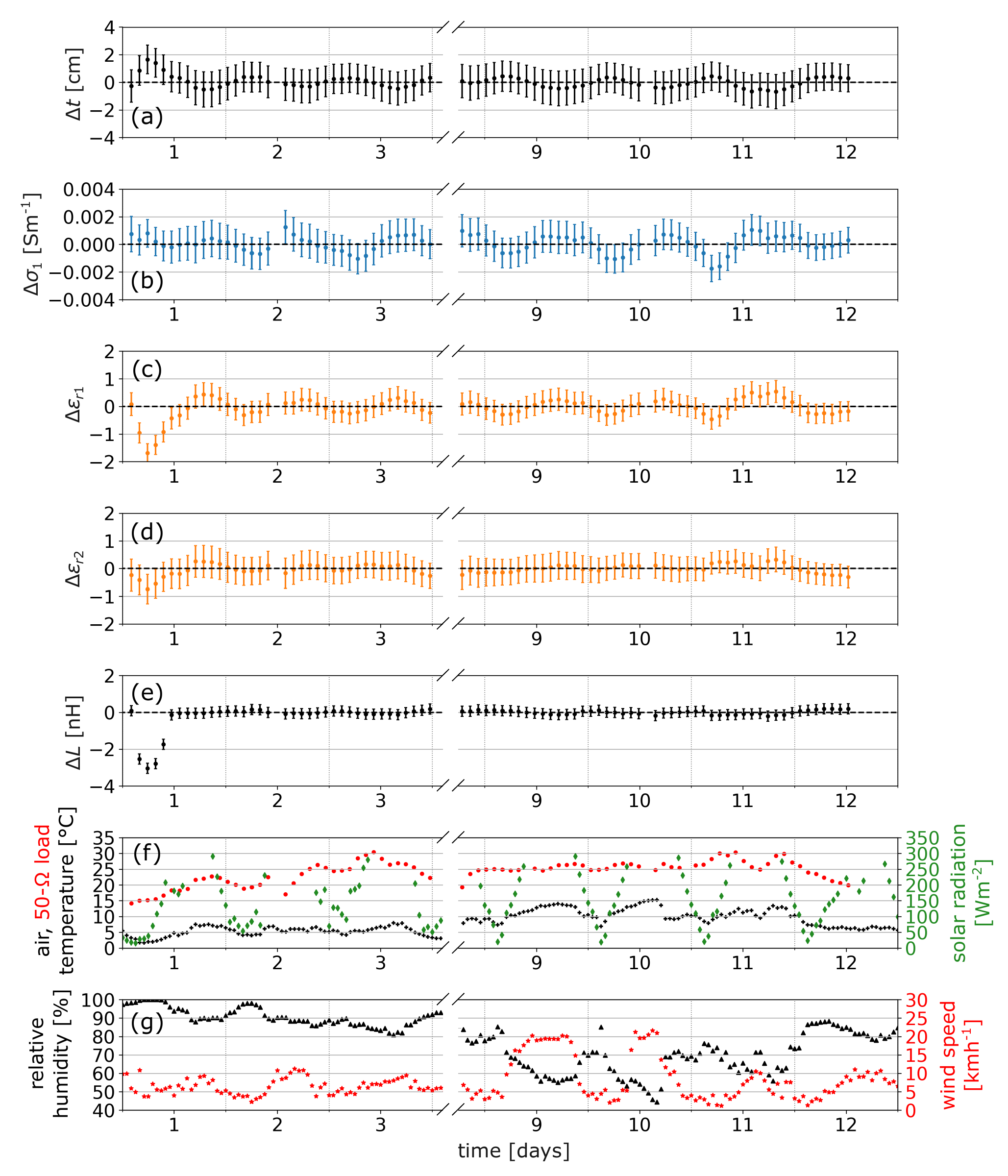}
\caption{Residuals after removing best-fit slopes from chunks one and two of the two-layer parameter estimates. When fitting slopes to chunk one, we do not include points $2$--$5$ of day~$1$ because they have an anomalous behavior. (a) Top-layer thickness. (b) Top-layer conductivity. (c) Top-layer permittivity. (d) Bottom-layer permittivity. In this figure we do not include the bottom-layer conductivity because in the paper we only report upper limits for this parameter. (e) Residual inductance. (f) Temperature of the internal $50$-$\Omega$ load used for impedance calibration, air temperature, and solar radiation. (g) Relative humidity and wind speed.}
\label{figure_sinusoidal_fluctuations}
\end{figure}

The supporting information includes text, figures, and tables that complement the analyses and findings described in the main manuscript. Text~S1 describes the calibration of the antenna impedance measurements. Text~S2 provides details of the EM simulations of the MIST instrument with Feko. Text~S3 describes the uncertainties accounted for in our analysis. Text~S4 provides details of the soil parameter fits. Figure~\ref{figure_map} shows a map of the study site. Figure~\ref{figure_instrument_dimensions} shows a diagram of the MIST instrument. Figure~\ref{figure_feko_simulations} shows all the Feko simulations done for the single- and two-layer soil models. Figure~\ref{figure_total_uncertainty} shows the uncertainties accounted for in our analysis. Figure~\ref{figure_sinusoidal_fluctuations} highlights the daily fluctuations of the soil parameter estimates after removing best-fit slopes from the estimates. Table~\ref{table_instrument_dimensions} lists the dimensions of the MIST instrument. Table~\ref{table_feko_simulations} provides details of the Feko simulations on the regular soil parameter grids. Table~\ref{table_results_sensitivity} presents the results obtained when fixing the residual inductance at zero and when constraining the top-layer thickness to the range of values obtained with metal probe measurements.

\clearpage

\noindent\textbf{Text S1. Calibration of the Antenna Impedance Measurements}

In this paper we estimate the soil parameters from the impedance calibrated at the excitation port of the antenna, which is located where the balun connects to the two aluminum panels. 
The raw antenna impedance measurements are performed by the vector network analyzer (VNA) located inside of the MIST receiver. These measurements are affected by unwanted signal contributions from the VNA itself, as well as from connectors, cables, switches, and balun that form the path between the VNA and the antenna excitation port \cite{monsalve2024}. These unwanted contributions are removed from each of the $85$ measurements in a calibration process conducted offline and consisting of three steps. In the first step, each raw measurement is calibrated at an abstract location, or ``plane'', inside the receiver, using the measurements from the internal OSL calibration standards belonging to the same $111$-min measurement block. The second step shifts the calibration plane to the receiver input using lab measurements of a different set of OSL standards externally connected to the input. The impedance calibrated in the second step represents the antenna including the effect of the balun. To shift the calibration plane to the excitation port and remove the effect of the balun, in the third calibration step we de-embed the S-parameters of the balun, which are measured in the lab, from the measurements calibrated in step two. The uncertainties in the calibrated measurements are described in Text~S3 and shown in Figure~\ref{figure_total_uncertainty}.

\clearpage

\noindent\textbf{Text S2. EM Simulations of the MIST Instrument with Feko}

The EM simulations are computed with the Feko (\url{https://altair.com/feko}) software and its method of moments (MoM) solver. The MoM solver is the best suited for radiation problems involving large and open geometries. Although the popular HFSS (\url{https://www.ansys.com/products/electronics/ansys-hfss}) and CST (\url{https://www.3ds.com/products/simulia/cst-studio-suite}) antenna simulation softwares also include MoM solvers, recent comparisons indicate that Feko's MoM solver is the most reliable for our purposes \cite{mahesh2021}. We compute the Feko simulations across the frequency range $25$--$125$~MHz. The computation time of the simulations roughly scales with number of frequency points requested. We request $101$ points linearly spaced across the range for a resolution of $1$~MHz, which is sufficient to capture the features of the measured impedance. The computation of each simulation takes about six minutes on our $64$-core AMD Ryzen Threadripper PRO 5995WX computer. 

The simulations are computed at soil parameter combinations arranged in regular soil parameter grids. The span, number of samples, and resolution used for each soil parameter in the simulation grids are listed in Table~\ref{table_feko_simulations}. Figure~\ref{figure_feko_simulations} shows the simulated impedances. The parameter spans and number of samples were chosen considering the tradeoff between the following objectives: (1) exploring the largest geophysically possible parameter space for our study site \cite{reynolds2011}; (2) achieving the highest possible grid sampling density and subsequent interpolation accuracy; and (3) minimizing the total computation time of the Feko simulations. For the single-layer model, we computed nine samples per parameter for a total of $81$~simulations, which took about eight hours on our computer. For the two-layer model, we computed four samples per parameter for a total of $1024$ simulations, which took about four days.

\clearpage

\noindent\textbf{Text S3. Measurement and Model Uncertainty}
\\
The total impedance uncertainty used in the soil parameter fits, $\delta$, is computed as the quadrature sum of the uncertainty in the impedance measurements, $\delta_{\textrm{meas}}$, and models, $\delta_{\textrm{model}}$. These two uncertainties, in turn, are computed as the quadrature sum of different source uncertainties. The source uncertainties are described below. Figure~\ref{figure_total_uncertainty} shows all the uncertainties at the $68\%$ confidence level. Text~S4 describes how $\delta$ is used in the fits.

\noindent \textbf{(1) Uncertainty in Impedance Measurements}

The uncertainty in the impedance measurements, $\delta_{\textrm{meas}}$, is computed as the quadrature sum of two source uncertainties. The first source uncertainty, $\delta_{\textrm{cal}}$, is the uncertainty in the calibration of the measured impedance at the input of the MIST receiver. This uncertainty is primarily due to potential errors in the modeling of the internal and external OSL calibration standards, as well as to effects from temperature fluctuations at the receiver input that are unaccounted for. The second source uncertainty, $\delta_{\textrm{balun}}$, is the uncertainty due to potential errors in the balun S-parameters used when shifting the calibration plane to the antenna excitation port. Both terms are estimated from a laboratory characterization.

\noindent \textbf{(2) Uncertainty in Impedance Models}

The uncertainty in the impedance models, $\delta_{\textrm{models}}$, is computed as the quadrature sum of three source uncertainties. The first source uncertainty, $\delta_{\textrm{Feko}}$, characterizes the intrinsic accuracy limits of Feko with the MoM solver when simulating the MIST instrument. We quantify this uncertainty by simulating in Feko a cylindrical transmission line and comparing the simulated impedance with the impedance predicted theoretically \cite{pozar2005}. When simulating the transmission line, we use the same Feko settings as when simulating the MIST instrument. The second source uncertainty, $\delta_{\textrm{geom}}$, accounts for potential errors in the geometry of the MIST instrument simulated in Feko compared with the true geometry in the field. To quantify this uncertainty, we run Feko simulations with realistic errors applied to the physical dimensions of the instrument, including the angles of the instrument relative to the soil. In the simulations, errors in different physical dimensions are applied one at a time. The geometry uncertainty is then computed as the quadrature sum of the differences between the perturbed simulated impedances and the nominal simulated impedance. The third source uncertainty, $\delta_{\textrm{int}}$, characterizes potential errors in the impedance interpolation. To quantify this uncertainty, we interpolate impedance models at random soil parameter combinations and compare these models with impedances directly simulated with Feko at the same parameter combinations. The interpolation uncertainty is then computed as the root mean square of the differences (RMSD) across models.

\clearpage

\noindent\textbf{Text S4. Fitting the Soil Parameters}

The soil parameters and their uncertainties are estimated using the \verb+curve_fit+ function from the \verb+SciPy+ Python package \cite{virtanen2020}. The \verb+curve_fit+ function implements a non-linear least squares algorithm that minimizes chi-squared, 

\begin{align}
\chi^2 = \sum_{i=1}^{N_d} \left(\frac{d_i-m_i}{\delta_i}\right)^2,
\label{equation_chi_squared}
\end{align}

\noindent where $d$ represents the impedance data, $m$ represents the impedance model, $\delta$ is the impedance uncertainty, and $N_d$ is the number of data points. In Equation~\ref{equation_chi_squared}, the uncertainty is incorporated at the $68\%$ confidence level and accounts for uncertainties in impedance measurements and models which, as described in Text~S3, are combined in quadrature. The uncertainty is shown in Figure~\ref{figure_total_uncertainty}. During each of the $85$ fits, the data, models, and uncertainties are provided to Equation~\ref{equation_chi_squared} as $202$-point arrays corresponding to the concatenation of the resistance and reactance of the antenna at $1$-MHz resolution over $25$--$125$~MHz. Therefore, $N_d=202$. In the fits, we impose geophysically realistic constraints on the soil parameters. Specifically, we constrain the parameters to ranges equal to the spans of the Feko simulations listed in Table~\ref{table_feko_simulations}. We also constrain $L$ to the very wide range $0$--$1$~H. We change the constraints only for the two tests listed in Table~\ref{table_results_sensitivity}. In one test we fix $L$ at zero and in the other we constrain $t$ to the range $50.9$--$53.5$~cm, which represents the thaw depth measurements done with the metal probe. The \verb+curve_fit+ function requires initial guesses. We tested many initial guesses within the ranges defined by the parameter constraints and confirmed that the results are insensitive to our choice. 

% \end{article}
\clearpage

\clearpage

\clearpage

\begin{table}
% \settablenum{S1}
\caption{Physical dimensions of the MIST instrument shown in Figure~\ref{figure_instrument_dimensions}.}
\label{table_instrument_dimensions}      
\centering                          
\begin{tabular}{l c c}        
\hline 
\\
Dimension & Parameter & Value [m] \\ 
\hline                        
antenna panel length       & al &  $1.2002$   \\
antenna panel width        & aw &  $0.5969$   \\ 
antenna panel thickness    & at &  $0.0032$ \\ 
antenna panel separation   & as &  $0.0215$  \\
antenna panel height       & ah &  $0.5220$   \\ 
receiver length         & rl &  $0.4050$ \\ 
receiver width          & rw &  $0.2600$ \\ 
receiver depth          & rd &  $0.3350$  \\
receiver height         & rh &  $0.0200$  \\
balun length         & bl &  $0.0500$  \\
balun width         & bw &  $0.0300$  \\
balun depth         & bd &  $0.0370$  \\
balun height         & bh &  $0.0220$  \\
\hline                                   
\end{tabular}
\end{table}

\begin{table}
\caption{Span, number of samples, and resolution used for each soil parameter in the Feko EM simulation grids.}
\label{table_feko_simulations}
\centering

\begin{tabular}{l l c c c}
\hline
Soil model & Parameter & Span & \# Samples & Resolution  \\
\hline
Single-layer & $\sigma$ &  0.0001--0.0281 Sm$^{-1}$ & 9 & 0.0093 Sm$^{-1}$ \\
& $\epsilon_{r}$ & 1--25 & 9 & 8 \\
\hline
Two-layer &  $t$ &  45--60 cm & 4 & 5 cm\\
& $\sigma_1$ & 0.002--0.026 Sm$^{-1}$ & 4 & 0.008 Sm$^{-1}$\\
& $\epsilon_{r1}$  & 1--25 & 4 & 8 \\
& $\sigma_2$ & 0--0.006 Sm$^{-1}$ & 4 & 0.002 Sm$^{-1}$\\
& $\epsilon_{r2}$  & 1--25 & 4 & 8 \\
\hline
\end{tabular}
\end{table}

% \begin{table}
% \caption{RMSD in parameter estimates when fixing $L$ at zero and constraining $t$ to $50.9$--$53.5$~cm. For reference, the third column shows the average of the $68\%$ parameter uncertainties from the nominal analysis.}
% \label{table_results_sensitivity}
% \centering
% \begin{tabular}{l l c c c}
% \hline
% Soil model & Parameter & Nominal $68\%$ uncertainty & RMSD, $L=0$~H & RMSD, $t=[50.9,53.5]$~cm \\
% \hline
% Single-layer & $\sigma$~[Sm$^{-1}$] & 0.0006 & 0.0007 & -- \\
% & $\epsilon_{r}$ & 0.2 & 0.2 & -- \\ 
% \hline
% Two-layer & $t$~[cm] & 1.1 & 0.43 & 0.48 \\
% & $\sigma_1$~[Sm$^{-1}$] & 0.001 & 0.002 & 0.0004 \\
% & $\epsilon_{r1}$ & 0.4 & 0.2 & 0.3 \\ 
% & $\sigma_2$~[Sm$^{-1}$] & $0.002$ & $3 \times 10^{-16}$ & 0.00004 \\
% & $\epsilon_{r2}$ & 0.5 & 0.5 & 0.2 \\
% & $L$~[nH] & 0.3 & -- & 0.03 \\
% \hline
% \end{tabular}
% \end{table}

\begin{table}
\caption{RMSD in two-layer parameter estimates when fixing $L$ at zero and constraining $t$ to $50.9$--$53.5$~cm. For reference, the second column shows the average $68\%$ parameter uncertainties from the nominal analysis.}
\label{table_results_sensitivity}
\centering
\begin{tabular}{l c c c}
\hline
Parameter & Nominal $68\%$ uncertainty & RMSD, $L=0$~H & RMSD, $t=[50.9,53.5]$~cm \\
\hline
$t$~[cm] & 1.1 & 0.43 & 0.48 \\
$\sigma_1$~[Sm$^{-1}$] & 0.001 & 0.002 & 0.0004 \\
$\epsilon_{r1}$ & 0.4 & 0.2 & 0.3 \\ 
$\sigma_2$~[Sm$^{-1}$] & $0.002$ & $3 \times 10^{-16}$ & 0.00004 \\
$\epsilon_{r2}$ & 0.5 & 0.5 & 0.2 \\
$L$~[nH] & 0.3 & -- & 0.03 \\
\hline
\end{tabular}
\end{table}

% \clearpage

% \bibliography{references}

% \end{document}

\end{document}